\documentclass[twocolumn]{aastex63}

\usepackage{amsmath}
\usepackage{amssymb}
\usepackage{rotating}

\usepackage[T1]{fontenc}
\usepackage[utf8]{inputenc}

\newcommand{\ra}[4]{${#1}^{\rm h}{#2}^{\rm m}{#3}\fs{#4}$}
\newcommand{\dec}[4]{${#1}\arcdeg{#2}\arcmin{#3}\farcs{#4}$}

\newcommand\tE{t_{\rm E}}

\newcommand\murel{\mu_{\rm rel}}

\newcommand\thetaE{\theta_{\rm E}}

\received{}
\revised{}
\accepted{}

\shorttitle{Adaptive-optics observations of free-floating planet candidates}
\shortauthors{P. Mr\'oz et al.}

\begin{document}

\title{Free-floating or wide-orbit? Keck adaptive-optics observations of free-floating planet candidates detected with gravitational microlensing}

\correspondingauthor{Przemek Mr\'oz}
\email{pmroz@astrouw.edu.pl}

\author[0000-0001-7016-1692]{Przemek Mr\'oz}
\affil{Astronomical Observatory, University of Warsaw, Al. Ujazdowskie 4, 00-478 Warszawa, Poland}

\author[0000-0002-6079-3335]{Makiko Ban}
\affil{Astronomical Observatory, University of Warsaw, Al. Ujazdowskie 4, 00-478 Warszawa, Poland}

\author{Pierlou Marty}
\affil{Facult\'e des Sciences, Aix-Marseille Universit\'e, Marseille, France}

\author[0000-0002-9245-6368]{Rados\l{}aw Poleski}
\affil{Astronomical Observatory, University of Warsaw, Al. Ujazdowskie 4, 00-478 Warszawa, Poland}

\begin{abstract}
Recent detections of extremely short-timescale microlensing events imply the existence of a large population of Earth- to Neptune-mass planets that appear to have no host stars. However, it is currently unknown whether these objects are truly free-floating planets or whether they are in wide orbits around a distant host star. Here, we present an analysis of high-resolution imaging observations of five free-floating planet candidates collected with the Keck telescope. If these candidates were actually wide-orbit planets, then the light of the host would appear at a separation of 40--60\,mas from the microlensing source star. No such stars are detected. We carry out injection and recovery simulations to estimate the sensitivity to putative host stars at different separations. Depending on the object, the presented observations rule out 11\%--36\% of potential hosts assuming that the probability of hosting a planet does not depend on the host mass. The results are sensitive to the latter assumption, and the probability of detecting the host star in the analyzed images may be a factor of $1.9 \pm 0.1$ larger, if the exoplanet hosting probability scales as the first power of the host star mass, as suggested by recent studies of planetary microlensing events. We argue that deeper observations, for example with \textit{JWST}, are needed to confidently confirm or refute the free-floating planet hypothesis. 
\end{abstract}


\section{Introduction} \label{sec:intro}

High-cadence observations of the Galactic bulge carried out by microlensing surveys in the past decade \cite[e.g.,][]{mroz2017, gould2022, sumi2023} have revealed a large population of Earth- to Neptune-mass planets that appear to have no host stars and may outnumber stars in the Milky Way. However, it is still unclear whether these objects are truly free-floating planets, untethered to any star, or whether they are in deceivingly wide orbits around a distant host star.

Evidence for that population was presented for the first time by \citet{mroz2017}, who carried out a systematic search for gravitational microlensing events in the data collected by the Optical Gravitational Lensing Experiment \citep[OGLE;][]{udalski2015} survey during the years 2010--2015. They selected 2617 microlensing events and found six events with extremely short timescales ($\tE = 0.1-0.5$\,days), which could not be explained by brown dwarfs or stars. Subsequent studies, carried out in collaboration between different surveys, including the Optical Gravitational Lensing Experiment (OGLE), the Korea Microlensing Telescope Network (KMTNet; \citealt{kim2016}), Microlensing Observations in Astrophysics (MOA; \citealt{bond2001}), and Wise \citep{yossi2016}, led to the measurement of angular Einstein radii $\thetaE$ and relative lens--source proper motions $\murel$ for several free-floating planet candidates \citep{mroz2018,mroz2019,mroz2020b,mroz2020d,kim2021,ryu2021}. These measurements ruled out an unusually high proper motion as a cause of short timescales of the detected events and strengthened the hypothesis that they originate from a population of Earth- to Neptune-mass planets.

\citet{gould2022}, following up on the idea presented by \citet{kim2021}, studied a sample of microlensing events with finite-source effects detected by KMTNet in 2016--2019. Interestingly, \citet{gould2022} reached the same conclusion as \citet{mroz2017} despite studying a different sample of events and using a different analysis method. \citet{gould2022} found two populations of microlensing events separated by a gap in Einstein radius, similar to the gap in the distribution of Einstein timescales near $\tE \sim 0.5$\,d that was detected by \citet{mroz2017}. They estimated that events located below the ``Einstein desert'' (in Einstein radii and timescales) are due to a power-law distribution of free-floating planet candidates with a mass function $dN/d\log{M} = (0.4 \pm 0.2) (M/38\,M_{\oplus})^{-p}$ per star with $0.9 \lesssim p \lesssim 1.2$. A similar conclusion was reached by \citet{sumi2023}, who studied a sample of microlensing events detected by the MOA survey in 2006--2014 \citep{koshimoto2023}. They inferred the normalization of $0.49^{+0.12}_{-0.32}$ and the slope of $p=0.94^{+0.47}_{-0.27}$ of the power-law mass function.

The light curves of short-timescale microlensing events detected in the OGLE, KMTNet, and MOA data are consistent with those of isolated objects. This does not necessarily mean that they are caused by free-floating planets. If the source passes sufficiently far from the planet's host star, then the host star would not leave any observable signatures in the microlensing light curve, making it virtually impossible to distinguish between free-floating and wide-orbit planet scenarios with only photometric observations of microlensing events \citep[see, e.g.,][]{han2003,han2005,clanton2017}. The likelihood of detecting the host decreases with the increasing host--planet separation. The available data enable us to rule out the presence of a putative host star within only approximately 10\,au of the planet \citep[e.g.,][]{mroz2018,mroz2020b,kim2021,ryu2021}.

It is therefore possible that some (or all) of the detected objects are actually in wide orbits around a distant host star, especially given the fact that we know very little about the population of wide-orbit exoplanets. The ice-giant planets in the solar system, Uranus and Neptune, orbit the Sun with orbital periods of 84 and 165 yr, respectively. Finding their analogs with transit and radial velocity methods is impossible in the foreseeable future \citep{kane2011}. However, wide-orbit planets may be quite common. Observations of nearby protoplanetary disks with the Atacama Large Millimeter/submillimeter Array \citep[ALMA; e.g.,][]{andrews2018} have revealed annular substructures that are believed to be the result of planet--disk interactions. Using these observations, \citet{zhang2018} estimated that about 50\% of the analyzed stars host a Neptune- to Jupiter-mass planet beyond 10\,au. \citet{poleski2021} conducted a systematic search for wide-orbit planets in the 20 yr of OGLE data and found that every microlensing star hosts $1.4^{+0.9}_{-0.6}$ ice-giant planets at separations from 5 to 15 au.
The origin of the wide-orbit planet population is still a subject of ongoing research. The ice-giant planets in the solar system, Uranus and Neptune, are believed to have formed inside the current orbit of Saturn and to have been scattered to their present orbits \citep[e.g.,][]{thommes1999,thommes2002,tsiganis2005}.

Thus, we are left with three possibilities: the short-timescale microlensing events that were detected by OGLE, KMTNet, and MOA may be caused by a population of ice-giant wide-orbit planets, free-floating planets, or some combination of both. Fortunately, there is a method that allows us to distinguish between these scenarios.

During the microlensing event, the lens and the source are nearly perfectly aligned and cannot be resolved. However, this situation is not static. The observer, lens, and source are moving relative to each other and the typical relative lens--source proper motion is on the order of 7\,mas\,yr$^{-1}$. Thus, several years after the event, the lens and the source should separate in the sky. If the lens has a host star, then the stellar light could be visible in the high-resolution images. If the lens is a free-floating planet, then no additional light source would be detected. 

High-resolution ground-based observations of microlensing events were previously used to derive precise masses of (bound) exoplanets \citep[e.g.,][]{beaulieu2016,beaulieu2018,bhattacharya2018,bhattacharya2021,bennett2020}, discover a planet orbiting a white dwarf \citep{blackman2021}, detect astrometric microlensing effects \citep{lu2016}, and verify microlensing models \citep[e.g.,][]{abdurrahman2021,shan2021}. However, no high-resolution observations of free-floating planet candidates have been reported in the literature. We obtained adaptive-optics observations of six short-timescale microlensing events that were identified by \citet{mroz2017}. In this paper, we describe the analysis of these images and derive the limits on the presence of putative host stars for five of the targets.

\begin{deluxetable*}{lcrrrrrrrr}[t]
\tabletypesize{\footnotesize}
\tablecaption{Free-floating planet candidates observed with Keck/NIRC2\label{tab:targets}}
\tablehead{
\colhead{Event} & \colhead{EWS Name} & \colhead{R.A.} & \colhead{Decl.} & \colhead{$\Delta t$} & \colhead{$n_{\rm wide}$} & \colhead{$n_{\rm narrow}$} & \colhead{$H$} & \colhead{$\alpha$} & \colhead{FWHM} \\
\colhead{} & \colhead{} & \colhead{} &\colhead{} & \colhead{(yr)} & \colhead{} &\colhead{} &\colhead{(mag)} & \colhead{} & \colhead{(mas)}}
\startdata
BLG501.26.33361  & OGLE-2011-BLG-0284 & \ra{17}{54}{17}{54} & \dec{-29}{18}{17}{0} & 10.1 &  3 & 0 & $15.979 \pm 0.020$ & \dots & \dots\\
BLG512.18.22725  & \dots              & \ra{18}{05}{25}{00} & \dec{-28}{28}{23}{9} &  9.0 &  3 & 4 & $17.209 \pm 0.025$ & $0.115 \pm 0.007$ & 54\\
BLG500.10.140417 & OGLE-2012-BLG-1073 & \ra{17}{53}{16}{89} & \dec{-28}{40}{51}{4} &  8.9 &  3 & 5 & $17.133 \pm 0.024$ & $0.173 \pm 0.005$ & 84\\
BLG501.31.5900   & OGLE-2012-BLG-1396 & \ra{17}{50}{42}{45} & \dec{-29}{24}{49}{7} &  8.7 & 10 & 5 & $13.873 \pm 0.013$ & $0.085 \pm 0.003$ & 67\\
BLG505.27.114211 & OGLE-2015-BLG-1044 & \ra{17}{59}{04}{18} & \dec{-28}{36}{51}{7} &  6.0 &  9 & 6 & $17.652 \pm 0.034$ & $0.092 \pm 0.005$ & 76\\
BLG501.02.127000 & OGLE-2015-BLG-1200 & \ra{17}{53}{13}{44} & \dec{-30}{18}{59}{6} &  6.0 &  8 & 3 & $15.935 \pm 0.024$ & $0.272 \pm 0.005$ & 71\\
\enddata
\tablecomments{Targets were taken from \citet{mroz2017}; we also provide the event name from the OGLE Early Warning System \citep[EWS;][]{udalski2003}, if available. The equatorial coordinates are for the J2000 equinox. $\Delta t$ is the time elapsed since the event, $n_{\rm wide}$ and $n_{\rm narrow}$ are the number of wide- and narrow-camera images, $H$ is the $H$-band magnitude of the microlensing source star, and $\alpha$ describes the accuracy of the PSF determination on narrow-camera images (Section~\ref{sec:fit}).}
\end{deluxetable*}

\section{Data and Calibrations}
\label{sec:data}

The observations were taken with the NIRC2 instrument behind the Laser Guide Star Adaptive Optics system mounted on the Keck II telescope \citep{wizinowich2006,vandam2006} at the W.~M.~Keck Observatory on Maunakea, Hawaii. The images were collected through the $H$-band filter with the wide and narrow cameras, which have plate scales of 0.04\,arcsec\,pixel$^{-1}$ and 0.01\,arcsec\,pixel$^{-1}$, respectively. The detector is a $1024 \times 1024$ Aladdin-3 InSb array.

Basic information about the targets is presented in Table~\ref{tab:targets}. All objects were observed during two nights, on 2021 June 2 and 2021 June 3, i.e., $\Delta t = 6 - 10$\,yr after the peak of the event. The observing conditions were suboptimal during both nights, with the seeing varying in the range $0.8-1.2''$. The exposure times were 60 and 180\,s for the wide and narrow camera, respectively. Three to ten (six) images were secured with the wide (narrow) camera for every target (except OGLE-2011-BLG-0284, for which no useful narrow-camera images were obtained because of a malfunction of the adaptive-optics system). The images were dithered to enable correction for bad pixels.

The images were corrected for dark current and flat frames using the \textsc{ccdproc} package \citep{craig2022} and were subsequently aligned and stacked using the \textsc{astroalign} package \citep{beroiz2020}. We used \textsc{SExtractor} \citep{bertin1996} in the MAG-AUTO mode to extract photometry from wide camera images. The instrumental magnitudes were calibrated against the Vista Variables in the Via L\'actea (VVV) source catalog \citep[][third data release]{saito2012}. Depending on the target, 29--64 isolated stars were cross-identified between the NIRC2 wide camera and VVV catalogs. The accuracy of the photometric calibrations varied from 0.013 to 0.021 mag.

The upper panels of Figure~\ref{fig:target1}--\ref{fig:target5} present the images of all of the targets analyzed. The upper left-hand panels show the $30'' \times 30''$ OGLE $I$-band reference image, the event is marked with a white cross. The upper middle panels show the NIRC2 wide camera $H$-band image of the same field, while the upper right-hand panels show the $5'' \times 5''$ NIRC2 narrow-camera zoom on the source star (which is also marked with a white cross). Figure~\ref{fig:target6} presents OGLE and NIRC2 wide camera images for OGLE-2011-BLG-0284. To identify the source in the NIRC2 images, we derived linear transformations between the OGLE reference frame and the Keck image coordinates. The calibrated $H$-band magnitudes of the sources are presented in Table~\ref{tab:targets}; the reported uncertainty is the sum in quadrature of the magnitude uncertainty reported by \textsc{SExtractor} and the calibration uncertainty.

\begin{figure*}
\includegraphics[width=0.5\textwidth]{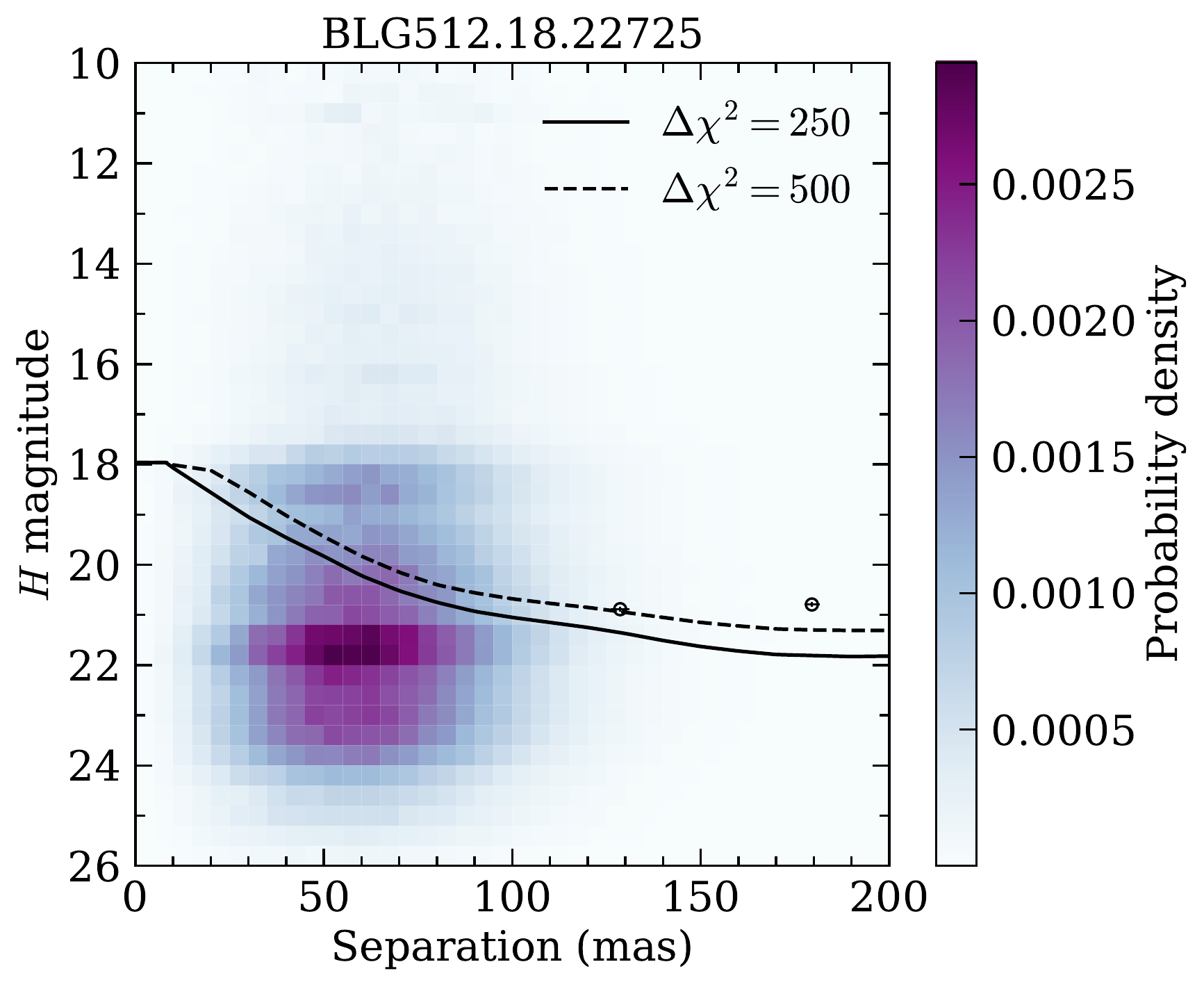}
\includegraphics[width=0.5\textwidth]{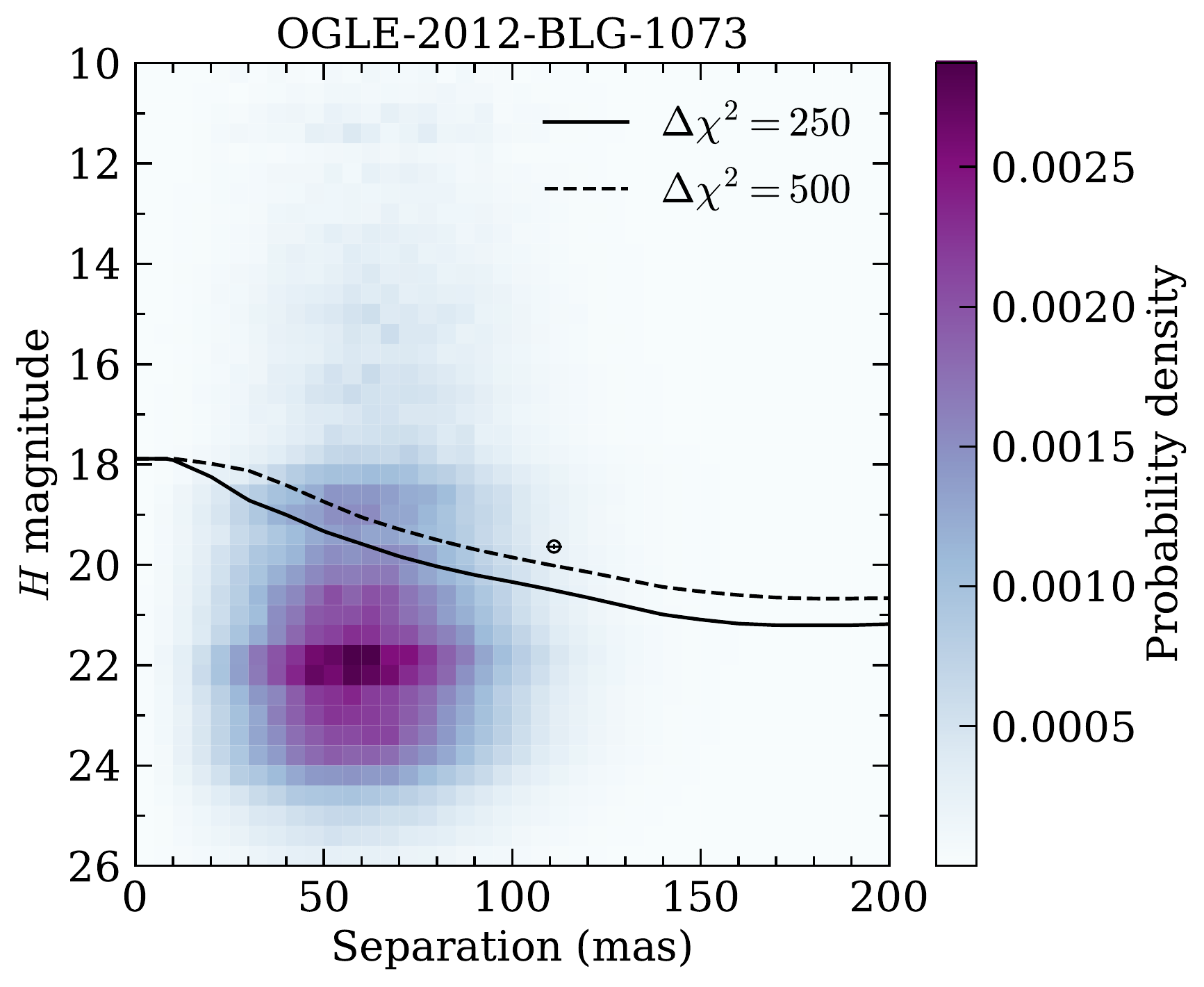} \\
\includegraphics[width=0.5\textwidth]{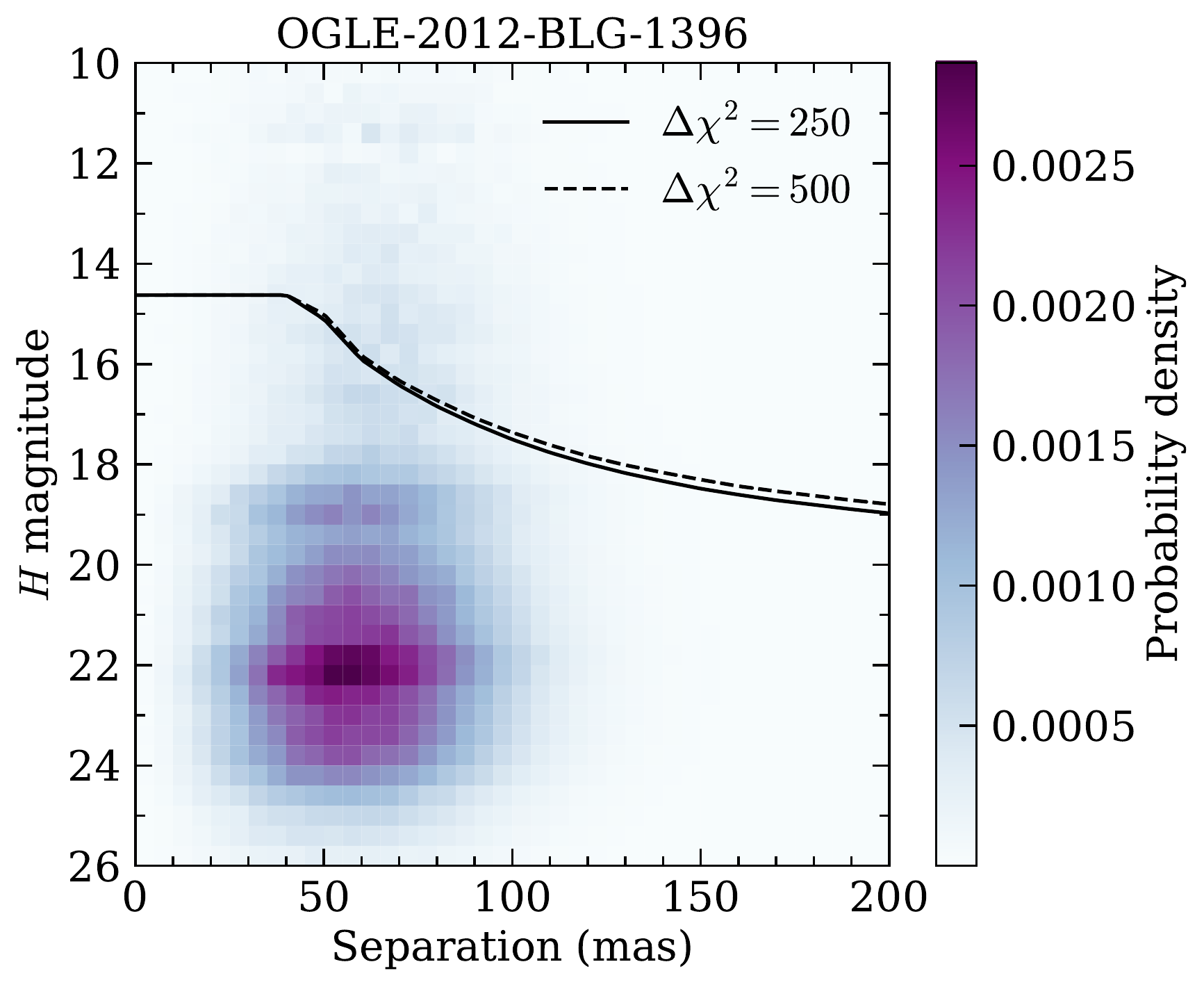}
\includegraphics[width=0.5\textwidth]{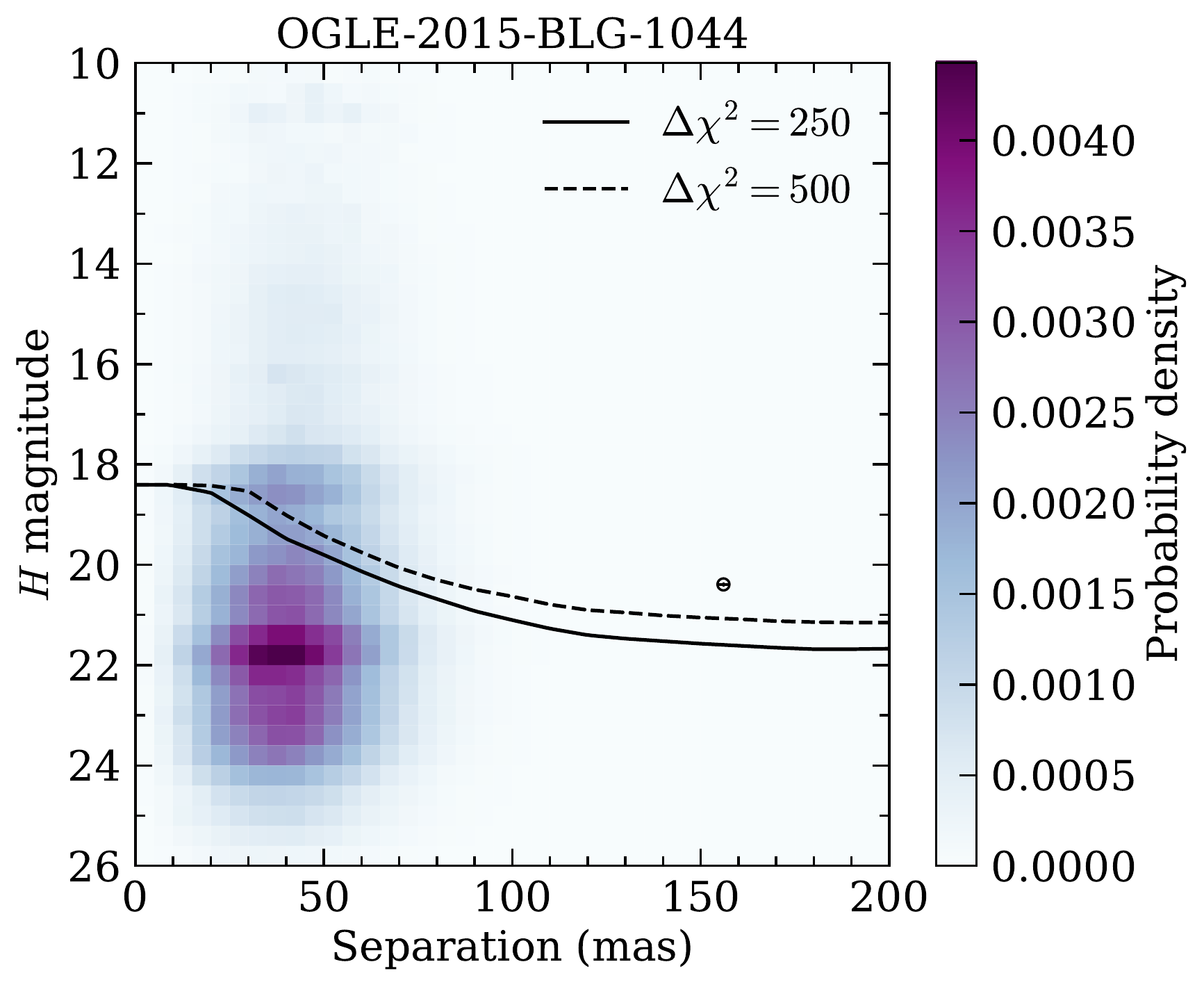} \\
\includegraphics[width=0.5\textwidth]{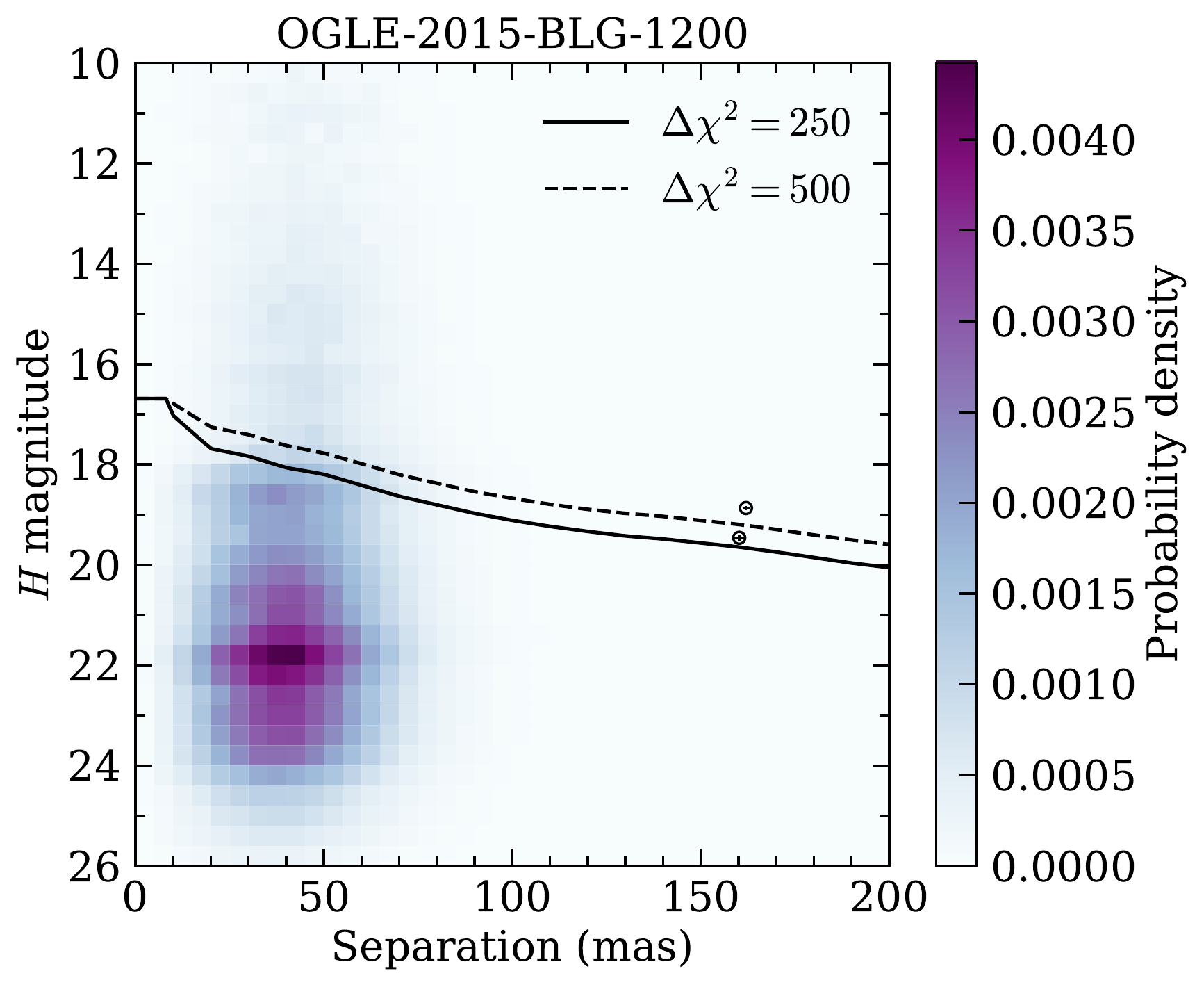}
\caption{Limits on putative host stars. Assuming that the analyzed objects are wide-orbit planets, the 2D histograms present the expected $H$-band brightness and lens--source separation of putative host stars during Keck observations. The black solid and dashed lines present the sensitivity of our observations (corresponding to $\Delta\chi^2_{\rm thresh}=250$ and $\Delta\chi^2_{\rm thresh}=500$, respectively). The black points mark the detected stars.}
\label{fig:limits}
\end{figure*}

\section{Searching for Putative Host Stars}
\label{sec:fit}

Once source stars were identified in the NIRC2 wide camera images, we used the narrow-camera images to search for putative host stars. We first constructed a model of the point spread function (PSF) for every target, following the algorithm devised by \citet{anderson2000}, as implemented in the \textsc{photutils} package \citep{bradley2022}. The PSF model was built using 5--24 (depending on the image) isolated, bright nearby stars. To minimize the effects of the spatial variation of the PSF, the selected stars were located within $5''$ of the target. We also made sure that the target itself was not used for the PSF construction. The PSF models are $25\times 25$ pixels in size, except for OGLE-2015-BLG-1200 (for which a $45 \times 45$ pixel model was built) and OGLE-2012-BLG-1396. The latter target is the brightest in our sample ($H=13.873 \pm 0.013$) and the models created with \textsc{photutils} were unable to accurately model the extended wings of the PSF. Instead, we used the only star of comparable brightness in the narrow-camera field of view (located at \ra{17}{50}{42}{77}, \dec{-29}{24}{50}{7}) as a proxy of the PSF of that target. The full width at half maximum (FWHM) of the PSF varies from 54 to 84\,mas, depending on the target (Table~\ref{tab:targets}).

Subsequently, we used the created PSFs to measure the positions and fluxes of all stars in a $100 \times 100$ pixel cutout centered around each target. This was achieved by minimizing the function:
\begin{equation}
\chi^2 = \sum_{i,j}\frac{\left(F_{ij}-\sum_{k=1}^{N}F_k\mathrm{PSF}_{ij}(x_k,y_k)-B\right)^2}{\sigma^2_{ij}+\alpha^2\sum_{k=1}^{N}F^2_k\mathrm{PSF}^2_{ij}(x_k,y_k)},
\label{eq:chi2}
\end{equation}
where $F_{ij}$ and $\sigma_{i,j}$ are the flux and flux uncertainty in the pixel $(i,j)$. Our model has $3N+1$ parameters: the flux $F_k$ and position $(x_k,y_k)$ of each star fitted ($k=1,\dots,N$, where $N$ is the number of stars fitted), and the mean background $B$. The parameter $\alpha$ describes the accuracy of the constructed PSFs: we assume that the uncertainty in the PSF shape can be parameterized as $\Delta\mathrm{PSF}=\alpha\mathrm{PSF}$. If $\alpha=0$, then we could treat the PSF as perfectly known. However, in the present case, the PSF is measured with some non-zero accuracy, especially given the fact that we noticed slight spatial variations of the PSF across the image.

In principle, we can treat $\alpha$ as a free parameter of the fit. However, the cutouts that we analyzed contain a relatively small number of stars (from two to seven), which are much fainter than the microlensing source star, and so information on $\alpha$ would mostly come from the shape of the profile of the source star itself. Therefore, we have to estimate $\alpha$ using an independent data set consisting of stars that are used to determine the shape of the PSF. 

To that end, we fitted single PSF models to all stars from that sample and calculated $\alpha$ that maximizes the following likelihood function:
\begin{align}
\begin{split}
\ln\mathcal{L} =& -\frac{1}{2}\sum_m\sum_{i,j}\frac{\left(F_{ij}-F_m\mathrm{PSF}_{ij}(x_m,y_m)-B_m\right)^2}{\sigma_{ij}^2+\alpha^2 F_m^2 \mathrm{PSF}^2_{ij}(x_m,y_m)}\\
& -\frac{1}{2}\sum_m\sum_{i,j}\ln\left(\sigma_{ij}^2+\alpha^2 F_m^2 \mathrm{PSF}^2_{ij}(x_m,y_m)\right).
\end{split}
\end{align}
Here, $(x_m,y_m)$, $F_m$, and $B_m$ denote the Cartesian coordinates, the total flux, and the background flux of the $m$th star from the sample used for the determination of PSF, respectively. We treat these parameters as nuisance parameters and marginalize over them to infer $\alpha$. The best-fit values are reported in Table~\ref{tab:targets}, they range from $0.085 \pm 0.003$ (for OGLE-2012-BLG-1396) to $0.272 \pm 0.005$ (for OGLE-2015-BLG-1200) with a median value of $0.115$. We checked that these numbers remain similar, or even slightly smaller, if we evaluate $\alpha$ using an independent set of isolated stars that were not used for the construction of the PSF.

Once we knew the accuracy of our PSF models, we fitted for the positions and fluxes of stars in the cutout images by minimizing the $\chi^2$ function (Eq.~\ref{eq:chi2}) using the Monte Carlo Markov chain (MCMC) algorithm coded by \citet{foreman2013}. We used flat (uniform) priors on all parameters. The initial positions $(x_k,y_k)$ were estimated using the Levenberg--Marquardt algorithm, while the initial fluxes $F_k$ and $B$ were calculated by solving a linear equation system derived by setting $\alpha=0$ in Eq.~\ref{eq:chi2}. 

Depending on the target, we modeled from two to seven stars. The residuals from the best-fit models are presented in the lower panels of Figures~\ref{fig:target1}--\ref{fig:target5}. The lower left-hand panels show the analyzed $100 \times 100$\,pixel image cutout with the microlensing source star in the center. The detected stars are marked with crosses and labeled. The lower middle panels show the same image after subtracting the source star, whereas the lower left-hand panels show the image after subtracting all detected stars. Positions, fluxes, and distances from the source star are presented in Tables~\ref{tab:target1} to \ref{tab:target5} in Appendix.

Residual images for BLG512.18.22725 and OGLE-2015-BLG-1044 are relatively smooth and appear not to indicate the presence of additional stars in the vicinity of the microlensing source star. However, the remaining subtractions are not perfect, most likely due to spatial variations of the PSF and inaccuracy of PSF models rather than the presence of additional stars, which is reflected by relatively large $\alpha=0.173$ (for OGLE-2012-BLG-1073) and $\alpha=0.272$ (for OGLE-2015-BLG-1200).

To test the hypothesis that additional stars are present near the source star, we carried out another round of MCMC modeling in which one additional star was added to the model, i.e., the model had three additional parameters. We specifically required that that star should be located in a $20 \times 20$ pixel box around the center of the image. We then calculated the quantity
\begin{equation}
\Delta\chi^2 \equiv \frac{\chi^2_{N}-\chi^2_{N+1}}{\chi^2_{N+1}/N_{\rm dof}},
\label{eq:dchi2}
\end{equation}
where $\chi^2_{N}$ and $\chi^2_{N+1}$ correspond to models with $N$ and $N+1$ stars, respectively, and $N_{\rm dof}$ is the number of degrees of freedom. We found that the $\Delta\chi^2$ improvement was modest, from 28 to 160. Furthermore, the residual images did not improve, e.g., a dipole pattern in the center of the lower right panel of Figure~\ref{fig:target5} could not be removed.

To check whether these detections were real or not, we carried out a series of image-level simulations (Section~\ref{sec:simuls}). We treated a star as detected if $\Delta\chi^2 > \Delta\chi^2_{\rm thresh}$ and, based on the results of our simulations, we adopted a cautiously large value of $\Delta\chi^2_{\rm thresh}=250$ because of the low-level correlated noise in the images. We also note that the faintest stars detected in the analyzed images (Tables~\ref{tab:target1}--\ref{tab:target5}) barely meet the adopted threshold (as shown in Figure~\ref{fig:limits}).

\section{Detection Efficiency Simulations}
\label{sec:simuls}

We carried out injection and recovery simulations to estimate the sensitivity of the images to possible host stars. We used the cutout images to create a $100 \times 100$ pixel artificial image in which we placed a synthetic star near the location of the microlensing source star. Subsequently, we fitted two models with $N$ and $N+1$ stars, respectively (where $N$ is the number of stars that were detected in the original image), by minimizing the $\chi^2$ function (Eq.~\ref{eq:chi2}). The positions of stars other than the microlensing source and the simulated synthetic star were fixed during the fitting, while the other parameters were allowed to vary. 

For every target, we created 7440 artificial images with the synthetic star placed at a separation of 10--200\,mas (with a step of 10\,mas) and a position angle of $30^{\circ}$ to $360^{\circ}$ (with a step of $30^{\circ}$) from the microlensing source star. We injected stars that are $\Delta H$ fainter than the microlensing source star with $\Delta H$ varying from 0 to 6\,mag with a step of 0.2\,mag and measured the $\Delta\chi^2$ improvement (Eq.~\ref{eq:dchi2})

For each position, we calculated the limiting magnitudes corresponding to $\Delta\chi^2_{\rm thresh}=250$ and $\Delta\chi^2_{\rm thresh}=500$ to check how sensitive the results are depending on the adopted detection threshold. We found that the limiting magnitudes weakly depend on the position angle. Thus, we averaged them over all angles. They are presented in Figure~\ref{fig:limits} with black solid and dashed lines.

The limiting magnitudes increase with increasing separation from the source star and reach $H=19-22$\,mag at 200\,mas, depending on the brightness of the source.

\begin{deluxetable*}{lrrrr}[t]
\tabletypesize{\footnotesize}
\tablecaption{Probability of Detecting the Host Star\label{tab:res}}
\tablehead{
\colhead{Event} & \multicolumn{2}{c}{$p$ ($n=0$) (default)} & \multicolumn{2}{c}{$p$ ($n=1$)} \\
\colhead{} & \colhead{$\Delta\chi^2_{\rm thresh}=250$} & \colhead{$\Delta\chi^2_{\rm thresh}=500$} & \colhead{$\Delta\chi^2_{\rm thresh}=250$} & \colhead{$\Delta\chi^2_{\rm thresh}=500$}
}
\startdata
BLG512.18.22725    & 0.362 & 0.318 & 0.642 & 0.590 \\
OGLE-2012-BLG-1073 & 0.303 & 0.253 & 0.578 & 0.507 \\
OGLE-2012-BLG-1396 & 0.110 & 0.107 & 0.200 & 0.192 \\
OGLE-2015-BLG-1044 & 0.295 & 0.257 & 0.565 & 0.513 \\
OGLE-2015-BLG-1200 & 0.173 & 0.151 & 0.375 & 0.334 \\
\enddata
\tablecomments{We assume that the probability of hosting a planet depends on the host mass $M_{\rm host}$ as $M^n_{\rm host}$. We adopt $n=0$ as the default value, i.e., the probability of hosting a planet does not depend on the host mass.}
\end{deluxetable*}

\section{Limits on Host Stars}
\label{sec:limits}

\begin{figure}
\includegraphics[width=.5\textwidth]{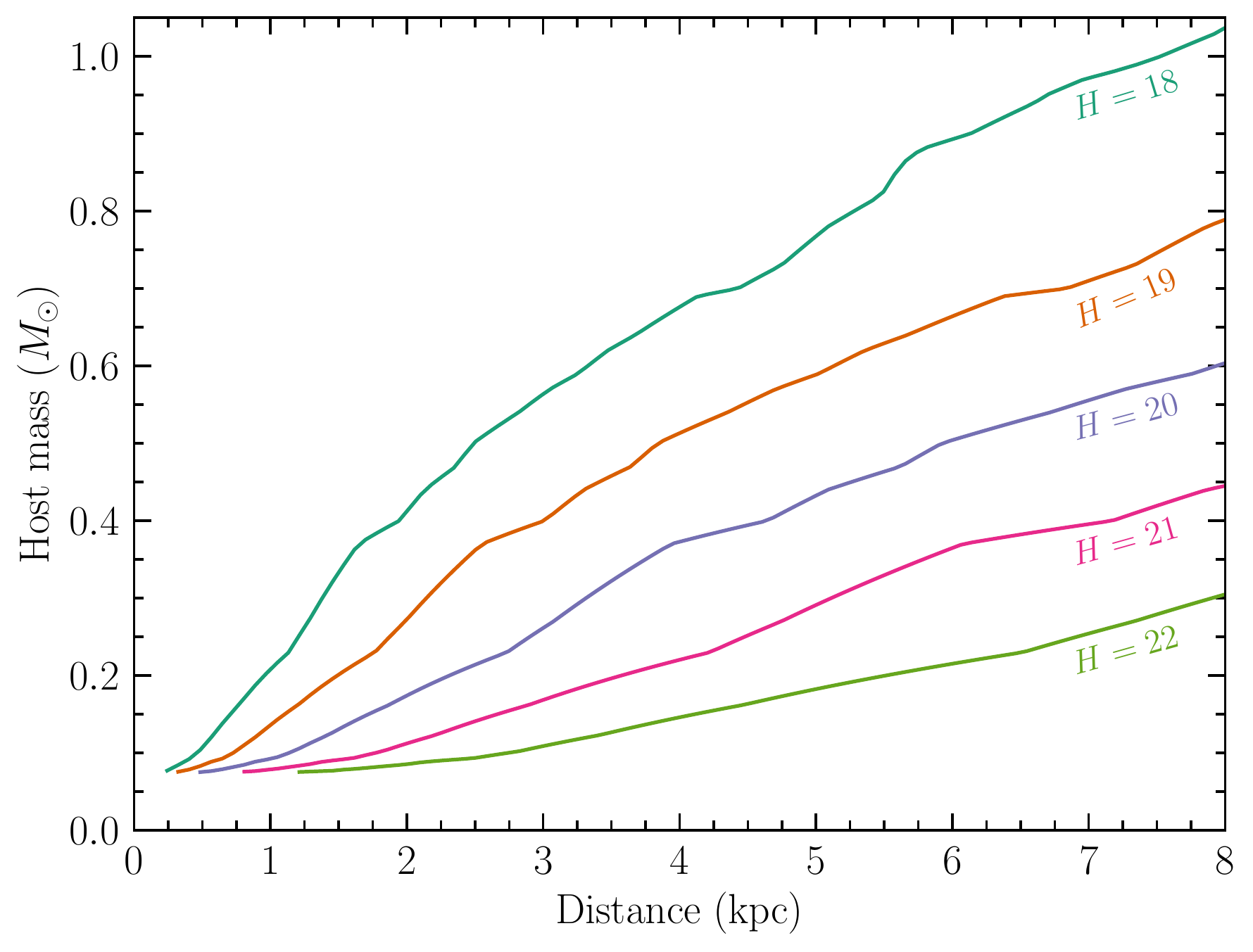}
\caption{The host mass--distance relations for a putative main-sequence host star of different apparent magnitude for the event OGLE-2015-BLG-1044.}
\label{fig:limits2}
\end{figure}

In this section, we assume that the analyzed objects are wide-orbit planets orbiting a distant host star and we calculate the probability of detecting the host star in the Keck images. We consider only main-sequence hosts. To illustrate the physical limits on host stars that we can obtain with Keck images, Figure~\ref{fig:limits2} shows the relation between mass and distance of a putative host star of OGLE-2015-BLG-1044 for five possible values of $H$-band apparent magnitude. The limiting magnitude of $H=18$\,mag allows us to rule out main-sequence host stars more massive than $\sim0.7\,M_{\odot}$ at 4\,kpc and $\sim 1\,M_{\odot}$ at 8\,kpc. For $H=22$\,mag, these limits are much deeper: they rule out main-sequence host stars more massive than $\sim0.15\,M_{\odot}$ at 4\,kpc and $\sim 0.3\,M_{\odot}$ at 8\,kpc.

The exact position of the host star (the lens) relative to the source star in the sky is a priori unknown. For some short-duration microlensing events, which exhibit finite-source effects, it is possible to measure the lens--source relative proper motion, and therefore calculate the expected lens--source separation. The position angle of the lens relative to the source can be calculated only if the microlensing parallax is measured, but such measurements are currently extremely challenging for short-timescale microlensing events. Nevertheless, because \citet{mroz2017} did not detect finite-source effects in the light curves of their free-floating planet candidates, the exact position of the host star cannot be predicted.

We therefore used a Milky Way model to predict the distribution of possible lens--source distances and $H$-band magnitudes of the host star. We used the Galactic model of \citet{batista2011}, which describes the spatial distribution and kinematics of lenses. We assumed that host stars follow the mass function of \citet{chabrier2003} and that the probability of hosting a planet does not depend on the mass of the lens.

Our model predicted the relative lens--source proper motion, which -- when propagated to the epoch of observations -- yielded the expected lens--source separation. To estimate the apparent magnitude of the lens, we used the empirical mass--absolute magnitude relation for main-sequence stars from \citet{pecaut2013} (we assumed that Galactic bulge stars more massive than $1\,M_{\odot}$ evolved off the main sequence and cannot be detected). We also assumed that extinction toward a given lens is proportional to the integrated density of interstellar matter (which follows the model of \citealt{sharma2011}), extinction is normalized using the extinction maps of \citet{gonzalez2012}, which are based on the VVV data.

Blue 2D histograms in Figure~\ref{fig:limits} present the expected probability density distribution for luminous lenses in the separation--magnitude diagram, separately for every target. The most likely expected lens--source separation varies from $\sim40$\,mas (for the most ``recent'' events from 2015) to $\sim 60$\,mas (for events that occurred in 2012). The most likely apparent brightness of the lens ranges between 18 and 24 mag. The peak at $H \approx 22$\,mag corresponds to typical $0.3\,M_{\odot}$ main-sequence stars located in the Galactic bulge, while the apparent overdensity at $H\approx 18.5$\,mag is caused by the flattening of the empirical mass--absolute magnitude relation near $0.8\,M_{\odot}$ \citep{pecaut2013}. The solid and dashed lines present our limiting magnitudes calculated using image-level simulations (Section~\ref{sec:simuls}). We consider the host star as detected if it is located above these lines. The probability that the host star can be detected in the Keck images is presented in Table~\ref{tab:res}. It ranges from 0.110 (for OGLE-2012-BLG-1396) to 0.362 (for BLG512.18.22725) for $\Delta\chi^2_{\rm thresh}=250$. Adopting a more conservative detection limit ($\Delta\chi^2_{\rm thresh}=500$), the probabilities of detecting host stars are slightly lower, 0.107 to 0.318. 

Our calculations are based on a crucial assumption that the probability of hosting a planet does not depend on the mass of the host star. Observations by radial velocity and direct imaging surveys indidate that more massive stars are more likely to host giant planets. \citet{johnson2010} found that the planet-hosting probability $P_{\rm host} \propto M_{\rm host}^n$, where $M_{\rm host}$ is the host mass and $n=1.0 \pm 0.3$, based on a sample of giant planets with semi-major axes smaller than 2.5\,au observed by the California Planet Survey. A newer study by \citet{fulton2021}, based on the California Legacy Survey data, corroborated the conclusion that the giant planet occurrence rate increases with stellar mass. A strong correlation between planet occurrence rate and host star mass (for giant planets with masses $2-13\,M_{\rm Jup}$ and semi-major axes of 3--100 au) was found by \citet{nielsen2019} using the Gemini Planet Imager Exoplanet Survey data, who inferred $n=2.0 \pm 1.0$. Meanwhile, obsevations by \textit{Kepler} spacecraft indicate that the occurrence of close Earth- to Neptune-sized planets is higher toward lower mass hosts \citep[e.g.,][]{howard2012,mulders2015}.

It is still unclear how the planet occurrence rate depends on the host mass in the semi-major axis range probed by gravitational microlensing surveys. \citet{koshimoto2021} found $n = 0.7^{+0.8}_{-0.6}$ assuming a flat prior on $0 \leq n \leq 2$, based on a comparison of the observed distribution of lens-source proper motions and those predicted by the Milky Way model. However, the negative values of $n$ were not excluded by the analysis presented by \citet{koshimoto2021}. A preliminary study of the adaptive-optics observations of exoplanets from the statistical sample of \citet{suzuki2016} indicate that $n$ is close to unity \citep{bennett2023}. If that is the case, then the probability of detecting the host in our data is of course higher. To check how our results are sensitive to the adopted value of $n$, we repeated our calculations assuming $n=1$, as indicated by the preliminary study reported by \citet{bennett2023}. The results are reported in the fourth and fifth columns of Table~\ref{tab:res}. The probability of detecting the host star in the analyzed images ranges from 0.20 to 0.64, and is on average $1.9 \pm 0.1$ higher than under the agnostic prior on $n=0$.

\section{Discussion and Conclusions}
\label{sec:conclusions}

In this paper, we analyzed high-resolution images of free-floating planet candidates detected by \citet{mroz2017}. We detected six stars at separations $110-200$\,mas from the microlensing source star. In particular, star \#5 in the field of OGLE-2012-BLG-1073 is located $111.07 \pm 2.05$\,mas from the source (Table~\ref{tab:target2}) and star \#6 in the field of BLG512.18.22725 is located $128.61 \pm 2.55$\,mas from the source (Table~\ref{tab:target1}).

Could any of the detected stars be a host star? We find it unlikely. According to our simulations, the probability that the lens would be at a separation greater than 110\,mas (130\,mas) and be brighter than $H=21$\,mag ($H=22$\,mag) is 0.031 (0.012) for OGLE-2012-BLG-1073 (BLG512.18.22725). Host stars located at separation greater than 150\,mas are even less probable ($3.6\times10^{-3}$ and $1.1\times 10^{-4}$ for events that occurred in 2012 and 2015, respectively).

We cannot completely rule out the possibility that some of the detected stars are actually hosts. This would require unlikely (but not impossibly) large relative lens--source proper motions greater than $\sim 12$\,mas\,yr$^{-1}$. This hypothesis can be tested by obtaining additional high-resolution images to measure the proper motion of those stars with respect to the source.

Going back to the question posed in the title, are free-floating planet candidates detected by microlensing surveys actually wide-orbit planets? The presented Keck images are not deep enough to confidently answer this question. Assuming that all events are indeed due to wide-orbit planets, the probability that no host stars were detected in the Keck images is $P=\prod_i(1-p_i)$ (where $p_i$ is the probability of detecting the host for the $i$th event). We find $P=0.23$ (assuming that the planet-hosting probability is independent of the host mass, i.e., $n=0$), which is not small enough to robustly reject the wide-orbit planet hypothesis. For comparison, even if we adopted $n=1$, we would find $P=0.033$, which would correspond to rejecting the wide-orbit planet hypothesis at only $2.1 \sigma$ level.

Deeper imaging observations, for example with the next-generation large ground-based telescopes \citep[e.g.,][]{gould2016,gould2023} or \textit{JWST}, are needed to confidently confirm or refute the free-floating planet hypothesis. For example, we simulated synthetic \textit{JWST} images using the Pandeia Engine software\footnote{https://pypi.org/project/pandeia.engine/} and the library of \textit{JWST} PSFs\footnote{https://stsci.app.box.com/v/jwst-simulated-psf-library} to derive the detection limits analogous to those presented in Figure~\ref{fig:limits}. Our simulations indicate that a $\sim 1800$\,s exposure with the \textit{JWST} Near Infrared Camera (NIRCam) would allow us to detect 66\%--87\% of potential host stars (assuming that every main-sequence star has the same probability of hosting a planet, irrespective of its mass). If no host stars were detected with \textit{JWST} observations, then the hypothesis that all targets are wide-orbit planets would be rejected with a 0.9997 probability, i.e., at the $\sim3.6\sigma$ level.

\section*{Acknowledgements}
We thank Calen Henderson for his help in the preparation of Keck observations and Katarzyna Gregorowicz for her initial work on Keck imaging data. Work by M.B. and R.P. was supported by Polish National Agency for Academic Exchange grant ``Polish Returns 2019'' to R.P.

The data presented herein were obtained at the W.~M.~Keck Observatory, which is operated as a scientific partnership among the California Institute of Technology, the University of California and the National Aeronautics and Space Administration. The Observatory was made possible by the generous financial support of the W. M. Keck Foundation. The authors wish to recognize and acknowledge the very significant cultural role and reverence that the summit of Maunakea has always had within the indigenous Hawaiian community.  We are most fortunate to have the opportunity to conduct observations from this mountain.

\facilities{Keck:II (NIRC2), OGLE}

\bibliographystyle{aasjournal}
\bibliography{pap}


\clearpage
\newpage

\appendix
\section{Appendix A}
\restartappendixnumbering

Here, we present the Keck imaging data for all targets analyzed in this study (Figures A1--A6), together with a detailed information about stars detected in the vicinity of the targets (Tables A1--A5).

\begin{figure}[!htbp]
\includegraphics[width=\textwidth]{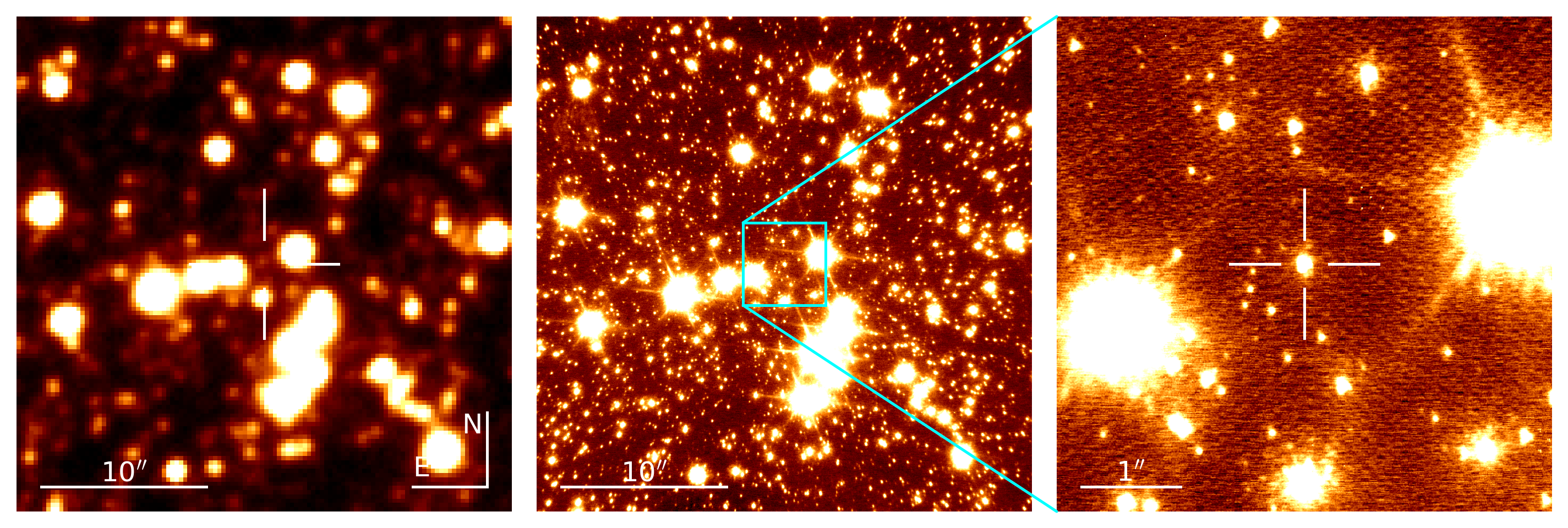}\\
\includegraphics[width=\textwidth]{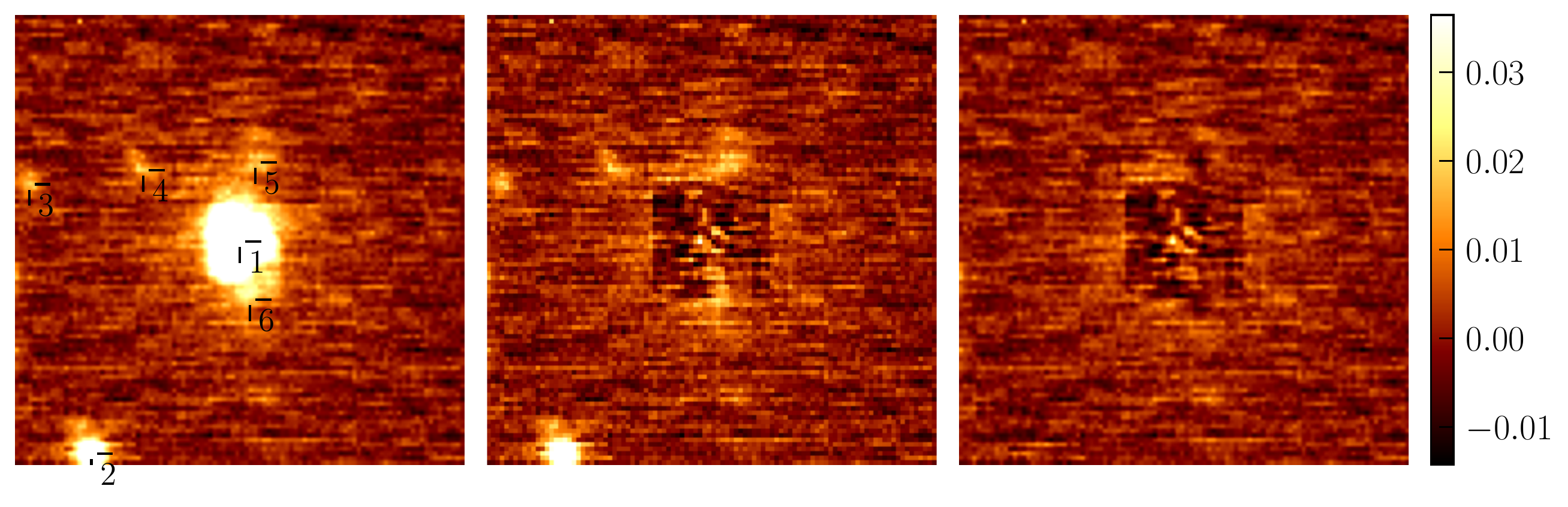}
\caption{Images for BLG512.18.22725. Upper panels: $30''\times30''$ OGLE $I$-band image (left-hand), $30''\times30''$ Keck NIRC2 wide camera $H$-band image (middle), and $5''\times 5''$ Keck NIRC2 narrow-camera $H$-band image (right-hand). Lower panels present a $1'' \times 1''$ zoom on the narrow-camera image. The lower middle panel presents the residuals after subtracting the microlensing source star, whereas the lower right-hand panel presents the residuals after subtracting all detected stars. The scale bar shows the flux renormalized to the peak of the target.}
\label{fig:target1}
\end{figure}

\clearpage
\newpage

\begin{deluxetable}{rrrrr}
\tablecaption{Stars detected in a $1'' \times 1''$ field around event BLG512.18.22725.\label{tab:target1}}
\tablehead{
\colhead{\#} & \colhead{$\Delta$E (mas)} & \colhead{$\Delta$N (mas)} & \colhead{$F/F_1$} & \colhead{Separation (mas)} }
\startdata
1 & $0$ & $0$ & $1$ & $0$ \\
2 & $ 328.99 \pm 0.52$ & $-468.84 \pm 0.59$ & $ 0.1104 \pm 0.0024$ & $572.76 \pm 0.57$ \\
3 & $ 466.82 \pm 1.96$ & $ 127.37 \pm 1.84$ & $ 0.0201 \pm 0.0012$ & $483.88 \pm 2.00$ \\
4 & $ 214.14 \pm 2.06$ & $ 158.64 \pm 1.75$ & $ 0.0246 \pm 0.0013$ & $266.49 \pm 2.14$ \\
5 & $ -33.80 \pm 2.31$ & $ 176.34 \pm 1.79$ & $ 0.0370 \pm 0.0017$ & $179.62 \pm 1.94$ \\
6 & $ -19.72 \pm 1.68$ & $-127.30 \pm 2.53$ & $ 0.0337 \pm 0.0018$ & $128.84 \pm 2.47$ \\
\enddata
\end{deluxetable}

\begin{figure}[!htbp]
\includegraphics[width=\textwidth]{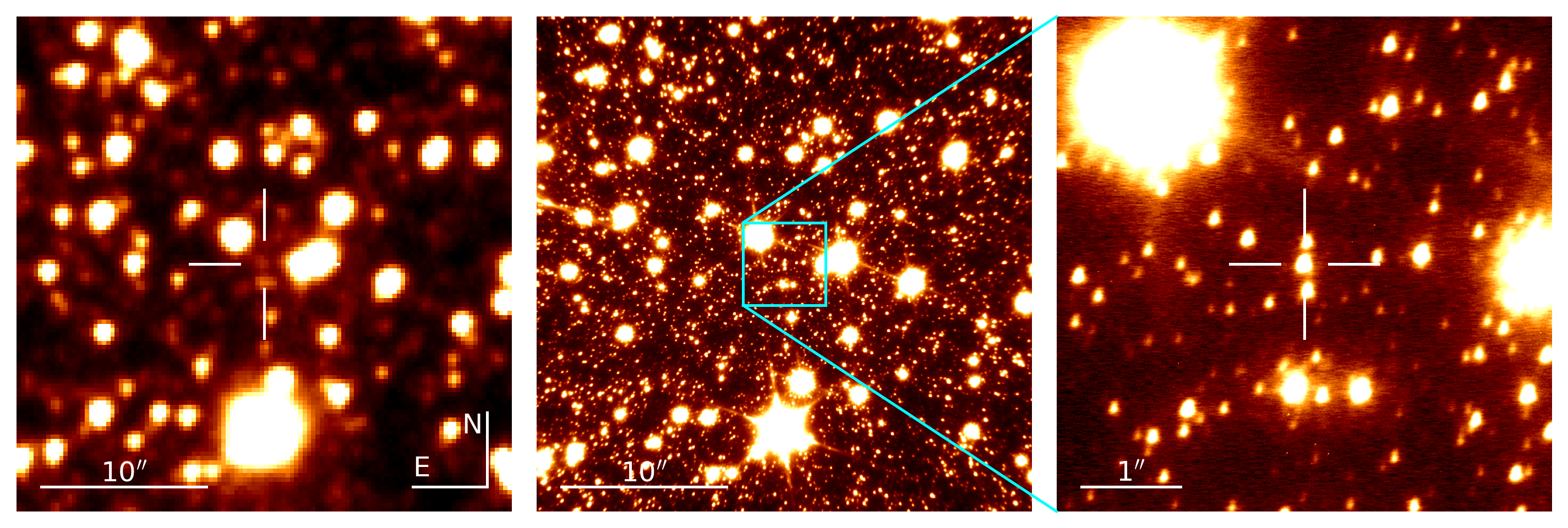}\\
\includegraphics[width=\textwidth]{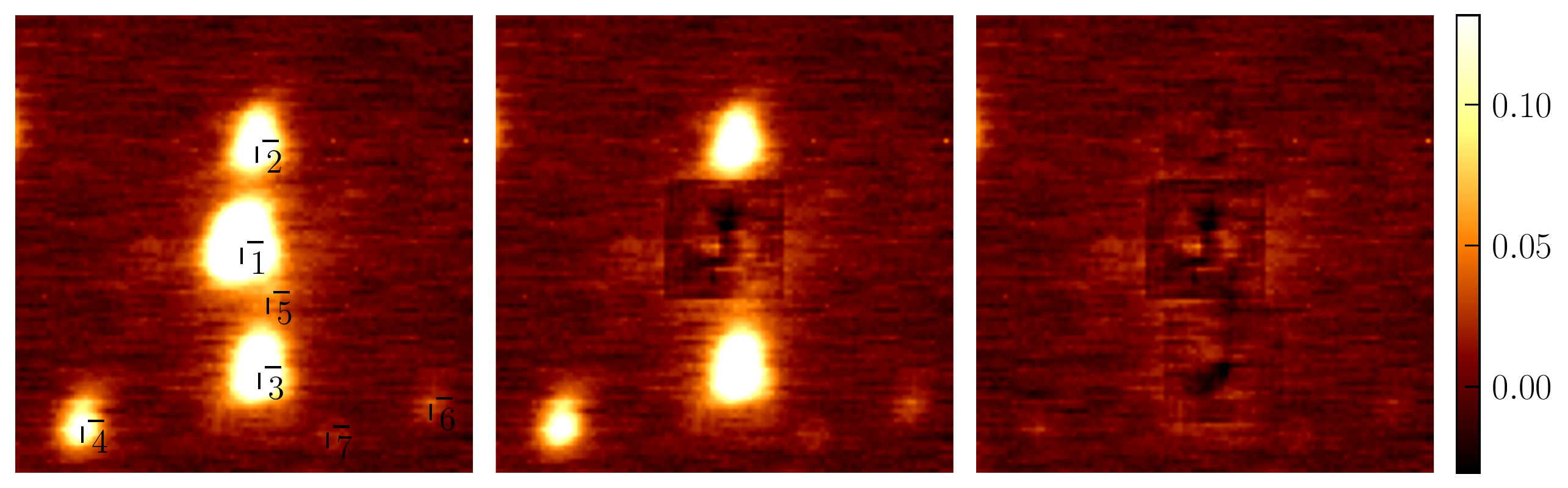}
\caption{Same as Figure~\ref{fig:target1}, but for event OGLE-2012-BLG-1073.}
\label{fig:target2}
\end{figure}

\clearpage
\newpage

\begin{deluxetable}{rrrrr}
\tablecaption{Stars detected in a $1'' \times 1''$ field around event OGLE-2012-BLG-1073.\label{tab:target2}}
\tablehead{
\colhead{\#} & \colhead{$\Delta$E (mas)} & \colhead{$\Delta$N (mas)} & \colhead{$F/F_1$} & \colhead{Separation (mas)} }
\startdata
1 & $0$ & $0$ & $1$ & $0$ \\
2 & $ -34.50 \pm 0.58$ & $ 219.72 \pm 0.55$ & $ 0.3938 \pm 0.0056$ & $222.42 \pm 0.57$ \\
3 & $ -36.67 \pm 0.53$ & $-272.39 \pm 0.68$ & $ 0.6166 \pm 0.0097$ & $274.84 \pm 0.65$ \\
4 & $ 344.35 \pm 0.65$ & $-388.94 \pm 0.77$ & $ 0.1927 \pm 0.0032$ & $519.46 \pm 0.74$ \\
5 & $ -66.77 \pm 2.02$ & $ -89.37 \pm 3.11$ & $ 0.1012 \pm 0.0048$ & $111.30 \pm 1.98$ \\
6 & $-412.16 \pm 1.74$ & $-340.44 \pm 2.71$ & $ 0.0294 \pm 0.0014$ & $534.55 \pm 1.69$ \\
7 & $-187.84 \pm 3.75$ & $-401.65 \pm 4.67$ & $ 0.0157 \pm 0.0014$ & $443.11 \pm 4.46$ \\
\enddata
\end{deluxetable}

\begin{figure}[!htbp]
\includegraphics[width=\textwidth]{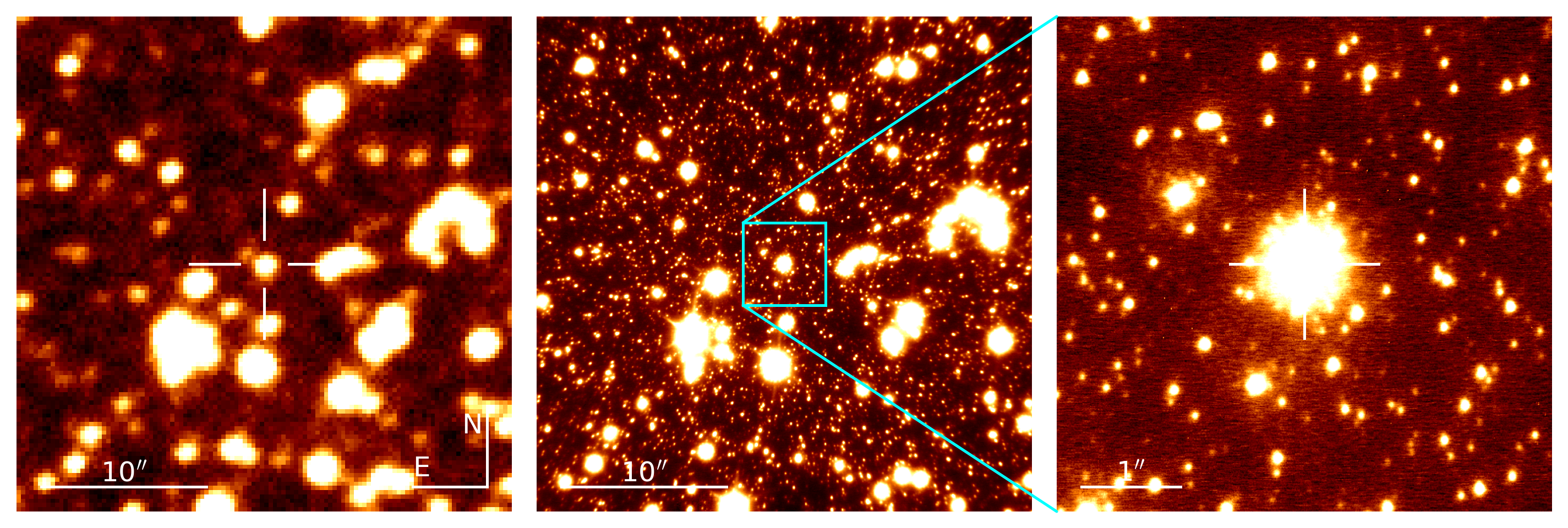}\\
\includegraphics[width=\textwidth]{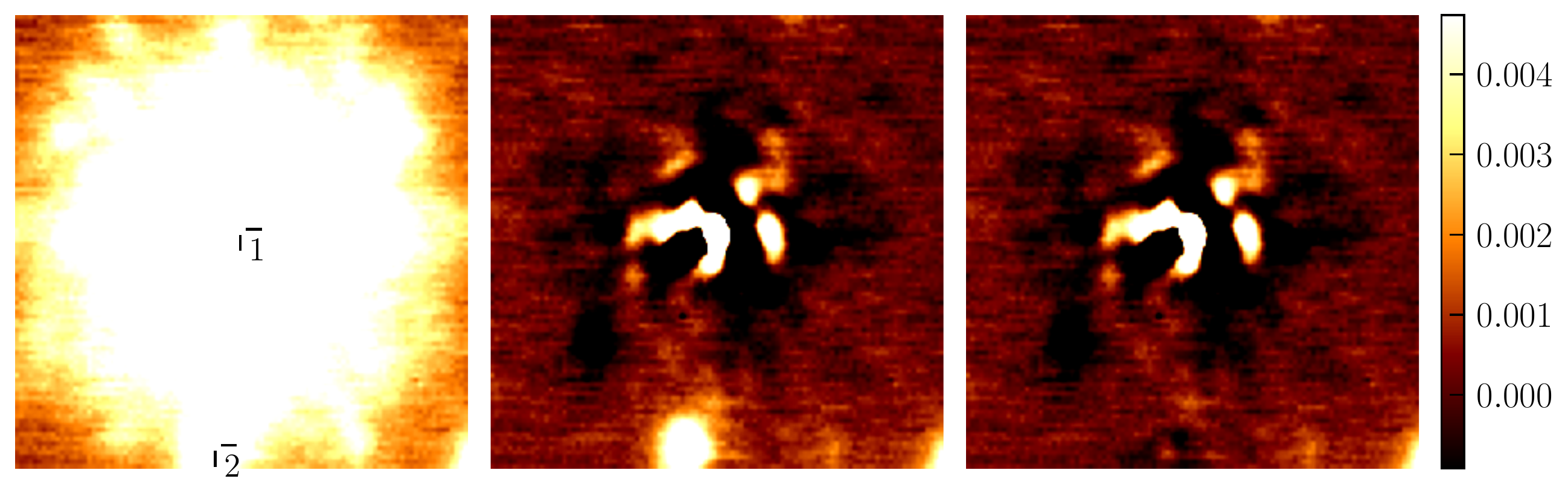}
\caption{Same as Figure~\ref{fig:target1}, but for event OGLE-2012-BLG-1396.}
\label{fig:target3}
\end{figure}

\clearpage
\newpage

\begin{deluxetable}{rrrrr}
\tablecaption{Stars detected in a $1'' \times 1''$ field around event OGLE-2012-BLG-1396.\label{tab:target3}}
\tablehead{
\colhead{\#} & \colhead{$\Delta$E (mas)} & \colhead{$\Delta$N (mas)} & \colhead{$F/F_1$} & \colhead{Separation (mas)} }
\startdata
1 & $0$ & $0$ & $1$ & $0$ \\
2 & $  54.26 \pm 0.43$ & $-472.97 \pm 0.49$ & $ 0.0208 \pm 0.0002$ & $476.07 \pm 0.49$ \\
\enddata
\end{deluxetable}

\begin{figure}[!htbp]
\includegraphics[width=\textwidth]{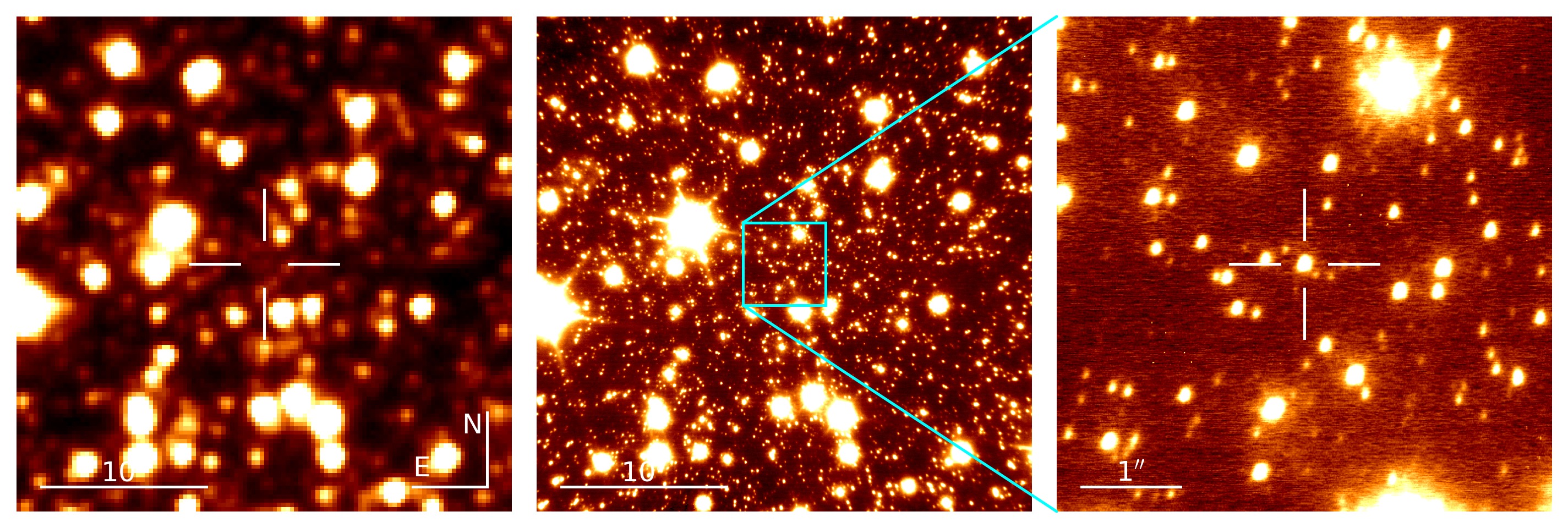}\\
\includegraphics[width=\textwidth]{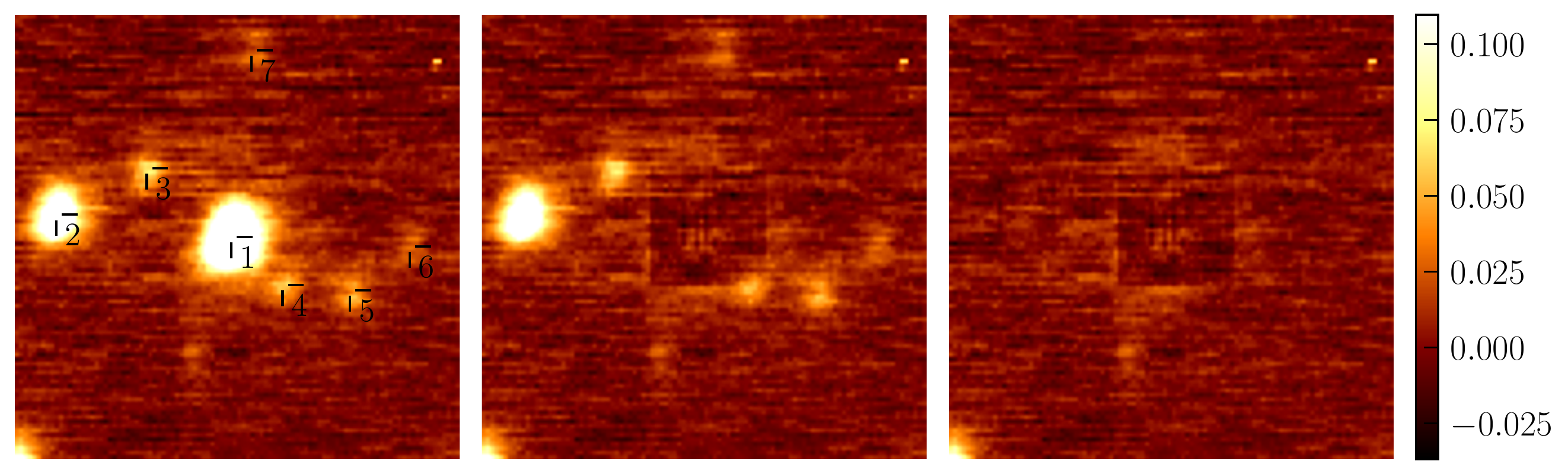}
\caption{Same as Figure~\ref{fig:target1}, but for event OGLE-2015-BLG-1044.}
\label{fig:target4}
\end{figure}

\clearpage
\newpage

\begin{deluxetable}{rrrrr}
\tablecaption{Stars detected in a $1'' \times 1''$ field around event OGLE-2015-BLG-1044.\label{tab:target4}}
\tablehead{
\colhead{\#} & \colhead{$\Delta$E (mas)} & \colhead{$\Delta$N (mas)} & \colhead{$F/F_1$} & \colhead{Separation (mas)} }
\startdata
1 & $0$ & $0$ & $1$ & $0$ \\
2 & $ 389.74 \pm 0.37$ & $  51.23 \pm 0.64$ & $ 0.4203 \pm 0.0105$ & $393.10 \pm 0.37$ \\
3 & $ 187.38 \pm 1.00$ & $ 155.21 \pm 2.15$ & $ 0.1042 \pm 0.0031$ & $243.06 \pm 1.01$ \\
4 & $-117.15 \pm 1.44$ & $-103.89 \pm 1.61$ & $ 0.0796 \pm 0.0025$ & $156.60 \pm 1.26$ \\
5 & $-267.86 \pm 1.17$ & $-118.24 \pm 1.62$ & $ 0.0797 \pm 0.0027$ & $292.72 \pm 1.17$ \\
6 & $-401.01 \pm 2.39$ & $ -19.88 \pm 3.58$ & $ 0.0416 \pm 0.0024$ & $401.50 \pm 2.40$ \\
7 & $ -46.21 \pm 2.05$ & $ 419.23 \pm 2.46$ & $ 0.0443 \pm 0.0024$ & $421.78 \pm 2.48$ \\
\enddata
\end{deluxetable}

\begin{figure}[!htbp]
\includegraphics[width=\textwidth]{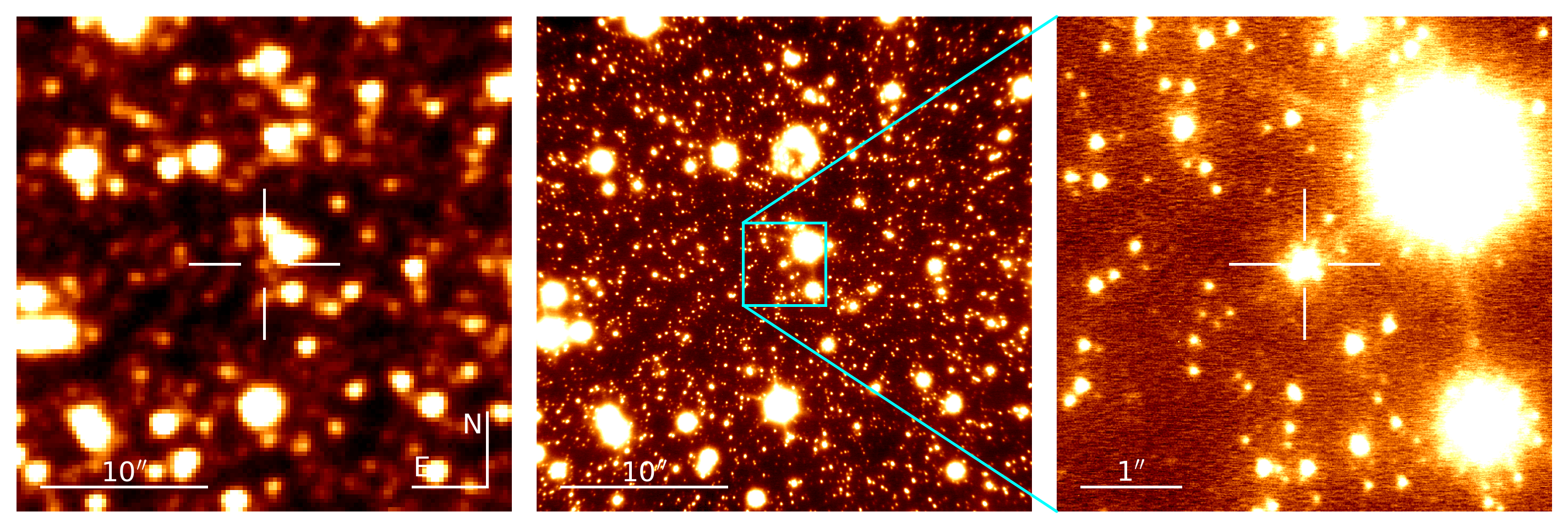}\\
\includegraphics[width=\textwidth]{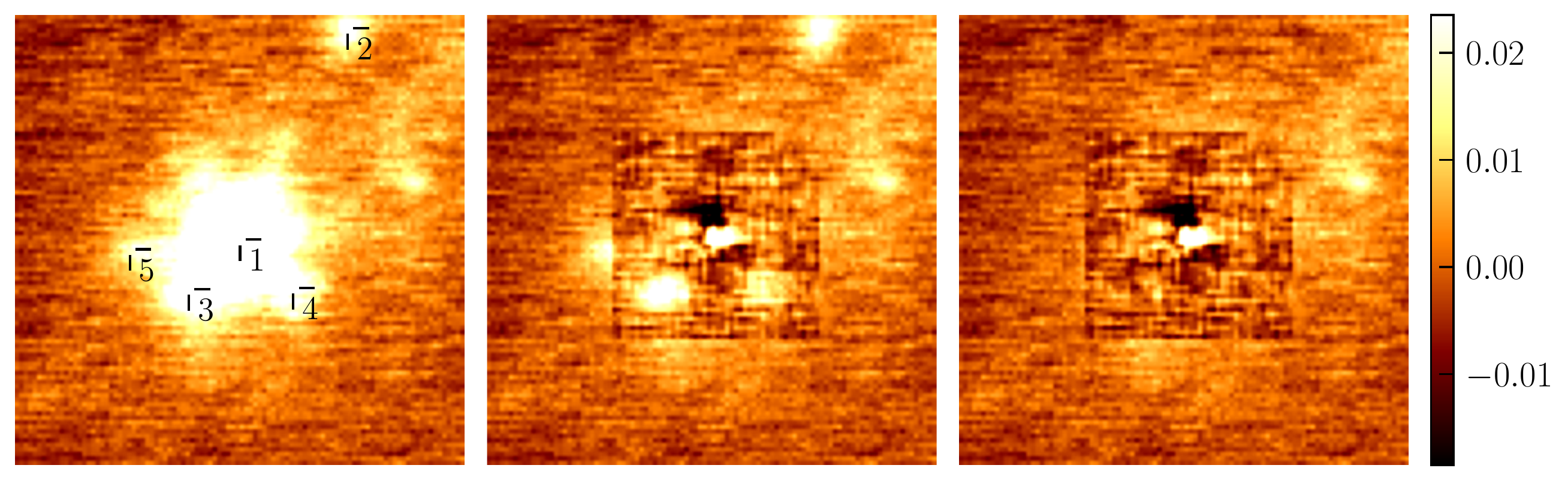}
\caption{Same as Figure~\ref{fig:target1}, but for event OGLE-2015-BLG-1200.}
\label{fig:target5}
\end{figure}

\clearpage
\newpage

\begin{deluxetable}{rrrrr}
\tablecaption{Stars detected in a $1'' \times 1''$ field around event OGLE-2015-BLG-1200.\label{tab:target5}}
\tablehead{
\colhead{\#} & \colhead{$\Delta$E (mas)} & \colhead{$\Delta$N (mas)} & \colhead{$F/F_1$} & \colhead{Separation (mas)} }
\startdata
1 & $0$ & $0$ & $1$ & $0$ \\
2 & $-243.98 \pm 0.65$ & $ 468.73 \pm 1.28$ & $ 0.0652 \pm 0.0025$ & $528.49 \pm 1.16$ \\
3 & $ 109.52 \pm 1.11$ & $-119.35 \pm 0.95$ & $ 0.0673 \pm 0.0029$ & $161.95 \pm 0.96$ \\
4 & $-118.74 \pm 1.89$ & $-106.91 \pm 3.11$ & $ 0.0390 \pm 0.0022$ & $159.80 \pm 2.49$ \\
5 & $ 242.43 \pm 1.43$ & $ -19.53 \pm 1.52$ & $ 0.0309 \pm 0.0012$ & $243.24 \pm 1.42$ \\
\enddata
\end{deluxetable}

\begin{figure}[!htbp]
\includegraphics[width=0.66\textwidth]{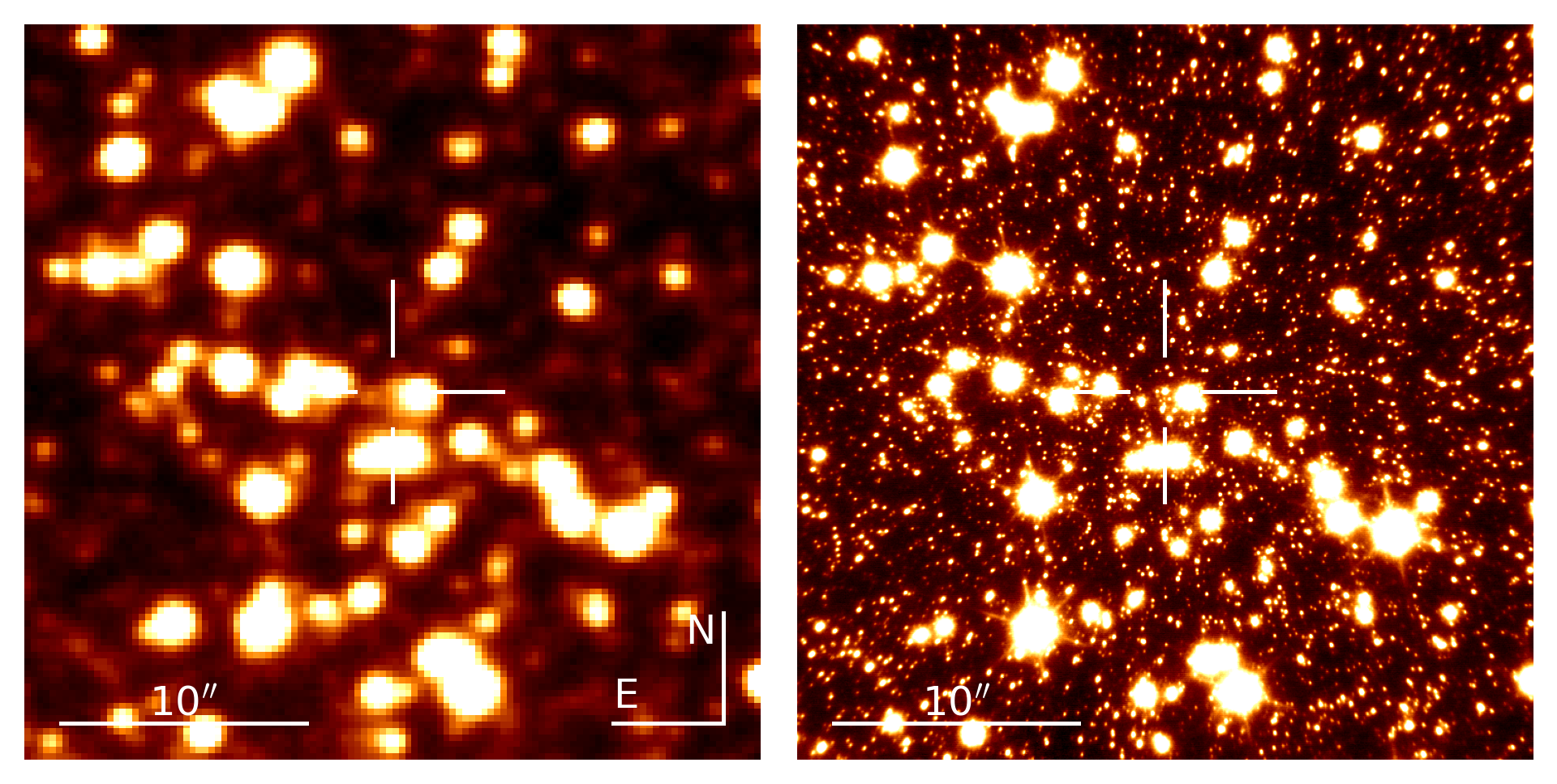}
\caption{Images for OGLE-2011-BLG-0284. Left-hand panel: $30''\times30''$ OGLE $I$-band image, right-hand panel: $30''\times30''$ Keck NIRC2 wide camera $H$-band image. No NIRC2 narrow-camera images of the target were obtained because of a malfunction of the adaptive-optics system.}
\label{fig:target6}
\end{figure}

\end{document}